\documentclass[aps,prl,twocolumn,showpacs]{revtex4}
\usepackage{amssymb}
\usepackage{amsmath}
\usepackage{graphicx}
\begin{document}
\title{Teleportation-Based Controlled-NOT Gate for Fault-Tolerant Quantum Computation}
\author{Alexander M. Goebel$^1$}
\author{Claudia Wagenknecht$^1$}
\author{Qiang Zhang$^{2}$}
\author{Yu-Ao Chen$^{1,2}$}
\author{Jian-Wei Pan$^{1,2}$}

\affiliation{$^1$Physikalisches Institut,
Ruprecht-Karls-Universit\"{a}t Heidelberg, Philosophenweg 12,
69120 Heidelberg, Germany\\
$^2$Hefei National Laboratory for Physical Sciences at Microscale
and Department of Modern Physics, University of Science and
Technology of China, Hefei, Anhui 230026, China}

\pacs{03.65.Ud, 03.67.Mn, 42.50.Dv, 42.50.Xa}

\begin{abstract}
Quantum computers promise dramatic speed ups for many computational
tasks. For large-scale quantum computation however, the inevitable
coupling of physical qubits to the noisy environment imposes a major
challenge for a real-life implementation. A scheme introduced by
Gottesmann and Chuang can help to overcome this difficulty by
performing universal quantum gates in a fault-tolerant manner. Here,
we report a non-trivial demonstration of this architecture by
performing a teleportation-based two-qubit controlled-NOT gate
through linear optics with a high-fidelity six-photon
interferometer. The obtained results clearly prove the involved
working principles and the entangling capability of the gate. Our
experiment represents an important step towards the feasibility of
realistic quantum computers and could trigger many further
applications in linear optics quantum information processing.
\end{abstract}

\date{\today}

\maketitle

In theory, quantum computers can outperform their classical
counterparts in various computational tasks such as searching an
unsorted data base \cite{Grover97Search} or factorizing large
numbers \cite{Shor94}. They further promise efficient simulation of
dynamics of complex quantum systems, which is not possible with
conventional computers \cite{Feynman82}. Practical implementations,
however suffer severely from coupling to the noisy environment and
residual imperfections in physical systems
\cite{Preskill98,Steane99,Gottesman98}.

Any system in nature couples to its environment. In quantum
computation this can lead to errors among the processed qubits
making quantum error correction schemes necessary. Several
algorithms to encode a logic qubit onto a number of physical qubits
have been developed
\cite{Shor95,Steane96,Calderbank96,Gottesman97misc,Bennett96EC,Laflamme96}.
These codes are able to correct for any single qubit error, as long
as maximally one of the physical qubits has been altered. After
decryption one is able to recover the unaltered, original logic
qubit. A next problem arises once we want to perform quantum gates,
i.e.~to perform logic operations on the protected data. Since the
logic qubit has been encoded, we need to perform corresponding
operations on the physical qubits. Depending on the characteristics
of the chosen code and gate (in particular conditional gates),
errors may then not only propagate between blocks of encoded qubits
but also within them. This can compromise the code's ability to
correct for these errors. The solution are the so-called
``fault-tolerant quantum gates''. A procedure is fault-tolerant if
its failing components (including the the errors in the encoded
input qubits) do not spread more errors in the block of encoded
output qubits than the code can correct.

In a seminal paper, Gottesman and Chuang introduced a novel protocol
to implement any quantum gate needed for quantum computation in a
fault-tolerant manner \cite{Gottesman99}. Their work has opened
doors to new ideas and has triggered several important protocols in
theoretical quantum information processing, such as one-way quantum
computation \cite{Raussendorf01} and linear optics quantum
computation \cite{KLM}. Although there is fast progress in the
theoretical description of quantum information processing, the
difficulties in handling quantum systems have not allowed an equal
advance in the experimental realization of the new proposals. Up to
now, not even a proof-in-principle demonstration of a
teleportation-based quantum logic gate, the fundamental building
block of the Gottesman-Chuang (GC) scheme, has been realized.

In this Letter, we report a non-trivial realization of the GC
scheme. We develop and exploit a high-fidelity six-photon
interferometer to combine the techniques of quantum teleportation
of a composite system \cite{Qiang06Teleportation} and the creation
of a four-qubit photon cluster state \cite{Kiesel05}. In the
experiment, we chose to implement a teleportation-based
controlled-NOT (C-NOT) gate, which, together with very easy to
implement single qubit operations, is sufficient to perform all
logic operations needed for quantum computation
\cite{Barenco95b,Nielsen04}. Compare to the previous six-photon
experiments \cite{Qiang06Teleportation,Goebel08} our experimental
setup is more complex and involves more interferences. Various
efforts have been made to achieve the stringent fidelity
requirements and sufficient six-photon count rate.

\begin{figure}[ptb]
\begin{center}
\includegraphics
[width=8cm]{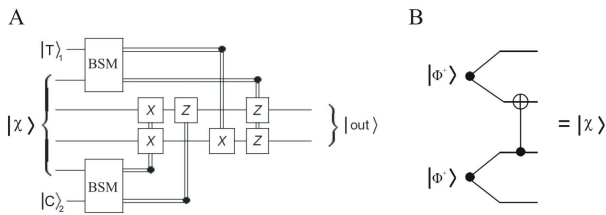}
\end{center}
\caption{Quantum circuit for teleporting two qubits through a C-NOT
gate. Time flow is from left to right. The input consisting of the
target qubit $|T\rangle_{1}$ and control qubit $|C\rangle_{2}$ can
be arbitrarily chosen. Bell State Measurements (BSMs) are performed
between the input states and the outer qubits of the special
entangled state $|\chi\rangle$. Depending on the outcome of the
BSMs, local unitary operations (X, Z) are conducted on the remaining
qubits of $|\chi\rangle$, which then form the output $|out\rangle =
U^{C-NOT} |T\rangle_{1}|C\rangle_{2}$. Single lines correspond to
qubits and double lines represent classical bits. \textbf{(B)} The
special entangle state $|\chi\rangle$ can be constructed by
performing a C-NOT gate on two EPR pairs, with $|\Phi^{+}\rangle =
\frac{1}{\sqrt{2}} \left(|H\rangle|H\rangle +
|V\rangle|V\rangle\right)$.} \label{fig1}
\end{figure}

The approach of Gottesman and Chuang, a generalization of quantum
teleportation \cite{Bennett93,Bouwmeester97}, is straight forward
and requires only a minimum of resources. A key element of their
work is the C-NOT gate, which acts on two qubits, a control and a
target qubit. The logic table of the C-NOT operation ($U^{C-NOT}$)
is given by $|H\rangle_{1}|H\rangle_{2} \rightarrow
|H\rangle_{1}|H\rangle_{2}$, $|H\rangle_{1}|V\rangle_{2} \rightarrow
|V\rangle_{1}|V\rangle_{2}$, $|V\rangle_{1}|H\rangle_{2} \rightarrow
|V\rangle_{1}|H\rangle_{2}$ and $|V\rangle_{1}|V\rangle_{2}
\rightarrow |H\rangle_{1}|V\rangle_{2}$, where we have used the
photon polarization degree of freedom to encode our qubits. A
schematic diagram of the procedure can be observed in
Fig.~\ref{fig1}A. One starts with the two input qubits $|T\rangle_1$
(target) and $|C\rangle_2$ (control). Instead of directly performing
complicated gate operations on the input qubits, one prepares in
forehand a special entangled four-qubit state $|\chi\rangle$. After
verification that the creation of $|\chi\rangle$ was successful, one
transfers the data of the input qubits onto $\vert\chi\rangle$ by
quantum teleportation. This is done by successively performing a
joint ``Bell-State-Measurement'' (BSM) between the target (control)
qubit and an outer qubit of $|\chi\rangle$, i.e. one projects the
target (control) qubit and one of the outer qubits of $|\chi\rangle$
onto a joint two-particle ``Bell state''. As a direct consequence of
the projective BSMs and the four-partite entanglement of
$|\chi\rangle$, the remaining two (output) qubits already posses the
information originally carried by the input qubits, i.e. the input
state is teleported onto the four-particle state $|\chi\rangle$. To
finish the procedure -- just like in the original teleportation
scheme -- we need to apply single qubit (Pauli) operations to the
output qubits, depending on the outcome of the BSMs.

Due to the special entanglement characteristics of $|\chi\rangle$,
the output state is equivalent to the desired unitary transformation
of the input state given by
\begin{equation}
|out\rangle= U^{C-NOT} |T\rangle_{1}|C\rangle_{2}.\label{out}
\end{equation}
This can be better understood by a closer look at the special
entangled state $|\chi\rangle$. It is a four-particle cluster state
\cite{Raussendorf03} of the form
\begin{eqnarray}
|\chi\rangle = \frac{1}{2} ( (|H\rangle|H\rangle +
|V\rangle|V\rangle) |H\rangle|H\rangle \nonumber\\ +
(|H\rangle|V\rangle + |V\rangle|H\rangle) |V\rangle|V\rangle).
\end{eqnarray}
which can be created simply by performing a C-NOT operation on two
EPR pairs as can be seen in Fig.~\ref{fig1}B. This C-NOT operation
is the essential difference to the original teleportation scheme and
is the reason for the fact that the output state is not identical to
the input state, but rather in the desired form of Eq.~\ref{out}. A
detailed discussion of the scheme is given in the supplementary
information.

Note, that in the above scheme all qubits are logic qubits. However,
the scheme generalizes in a straight forward manner when we use a
larger number of physical qubits to encode our logic qubits. The
procedure is then fault-tolerant since all operations are
transversal, i.e. qubits of one block of encoded qubits interact
only with corresponding qubits in other code blocks. A further
advantage is the fact that only classically controlled single-qubit
operations and BSMs are needed to perform the actual gate. The
resource of the special entangled state $|\chi\rangle$ can be
constructed in forehand. If its generation fails nothing is lost by
discarding it and trying again until successful generation. We would
like to emphasize two aspects: First, the setup can be used to
process any unknown input state and second, several other quantum
gates can be implemented by this scheme. The choice of gate only
depends on the form of the ancillary state $|\chi\rangle$.

\begin{figure*}[ptb]
\begin{center}
\includegraphics
[width=14cm]{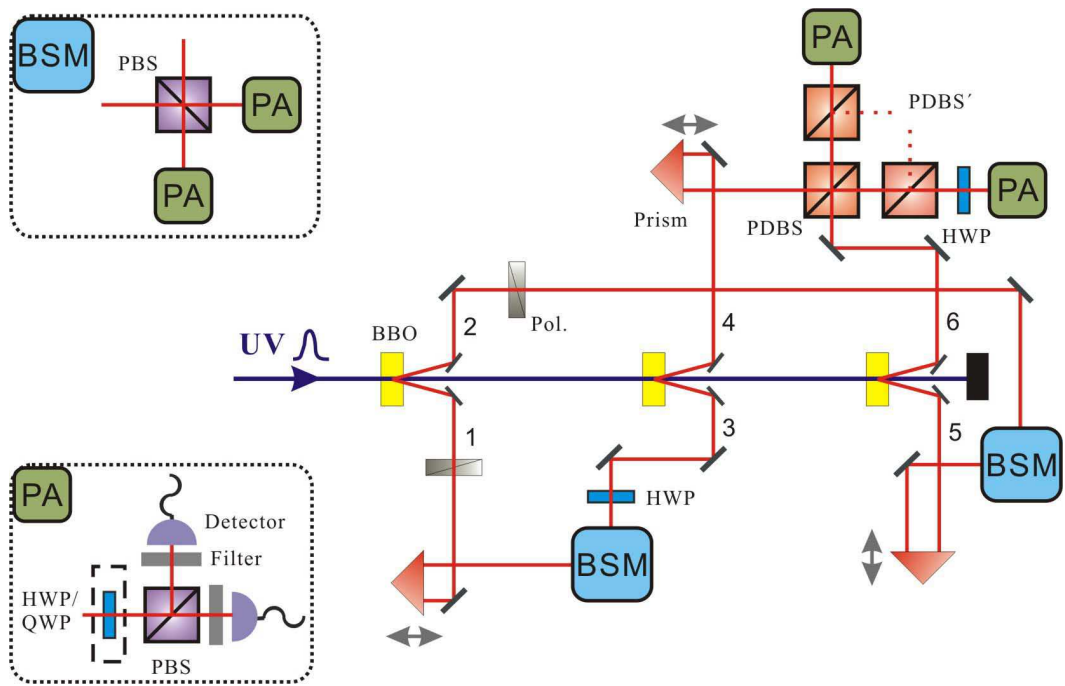}
\end{center}
\caption{A schematic diagram of the experimental setup. A
high-intensity pulsed ultraviolet laser beam (UV) at a central
wavelength of 390 nm, a pulse duration of 180 fs, and a repetition
rate of 76 MHz successively passes through three $\beta$-barium
borate (BBO) crystals to generate three polarization entangled
photon pairs via type-II spontaneous parametric down-conversion
\cite{Kwiat95}. At the first BBO the UV generates a photon pair in
modes 1 and 2 (i.e. the input consisting of the target and control
qubit). After the crystal, the UV is refocused onto the second BBO
to produce another entangled photon pair in modes 3 and 4 and
correspondingly for modes 5 and 6. Photons 4 and 6 are then
overlapped at a PDBS and together with photons 3 and 5 constitute
the cluster state. Two PDBS' are used for state normalization. The
prisms are mounted on step motors and are used to compensate the
time delay for the interference at the PDBS and the BSMs. A BSM is
performed by overlapping two incoming photons on a PBS and two
subsequent polarization analyses (PA). A PA projects the photon onto
an unambiguous polarization depending on the basis determined by the
choice of half or quarter wave plate (HWP or QWP). The photons are
detected by silicon avalanched single-photon detectors. Coincidences
are recorded with a coincidence unit clocked by the infrared laser
pulses. Pol. are polarizers to prepare the input state and Filter
label the narrow band filters with $\Delta_{FWHM}=3.2$ nm.}
\label{fig2}
\end{figure*}

A schematic diagram of our experimental setup is shown in
Fig.~\ref{fig2}. All three photon pairs are originally prepared in
the Bell-state $|\Phi^{+}\rangle =
\frac{1}{\sqrt{2}}(|H\rangle|H\rangle + |V\rangle|V\rangle)$. We
observe on average $7\times10^4$ photon pairs per second from each
(EPR) source. With this high-intensity entangled photon source we
obtain in total 3.5 six-photon events per minute. This is less
than half the count rate of our previous six-photon experiments
\cite{Qiang06Teleportation,Chaoyang07,Goebel08}. Since the new
scheme is more complex and involves more interferences, the
fidelity requirements are more stringent. Thus, we have to reduce
the pump power from 1.0 W to 0.8 W in order to reduce noise
contributions that arise from the emission of two pairs of
down-converted photons by a single source (double-pair-emission).

With the help of wave plates and polarizers, we prepare photon pair
1\&2 in the desired two-qubit input state $|\Psi\rangle_{12}$.
Photon pairs 3\&4 and 5\&6, which are both in the state
$|\Phi^{+}\rangle$, are used as a resource to construct the special
entangled state $|\chi\rangle_{3456}$.

Among the existing various methods for preparing the four-photon
cluster state $|\chi\rangle_{3456}$, however, only \cite{Kiesel05}
works in our present experiment as others are vulnerable to
double-pair-emission. As shown in Fig.~\ref{fig2} photons 4 and 6
are interfered on a beam splitter with a polarization-dependent
splitting ratio (PDBS), i.e. the transmission for horizontal
(vertical) polarization is $T_H=1$ ($T_V=1/3$). In order to
balance the transmission for all input polarizations, beam
splitters (PDBS') with reversed transmission conditions ($T_H =
1/3$, $T_V = 1$) are placed in each output of the overlapping
PDBS. Altogether, the probability of having one photon in each
desired output, and thus having successfully created the cluster
state, is $1/9$. Half wave plates (HWPs) in arms 3 and 4 are used
to transform the cluster state to the desired state by local
unitary operations.

With our high power EPR-source we are able to achieve a count rate
for the four-qubit cluster state $|\chi\rangle_{3456}$ that is two
orders of magnitude larger than in a recent experiment
\cite{Kiesel05}. We have measured the fidelity of
$|\chi\rangle_{3456}$ and obtain an experimental result of $0.694
\pm0.003$, which is only slightly lower than in \cite{Kiesel05} due
to the much higher pump power. The fidelity measurement has been
performed in complete analogy to Kiesel et al. \cite{Kiesel05}. The
improvement of the count rate is necessary in order to be able to
perform the six-photon experiment in a reasonable amount of time
over which the experimental setup can be kept stable.

Teleporting the input data of $|\psi\rangle_{12}$ to
$|\chi\rangle_{3456}$ requires joint BSMs on photons 1\&3 and
photons 2\&5. To demonstrate the working principle of the
teleportation-based C-NOT gate, it is sufficient to identify one
of the four Bell states in both BSMs
\cite{Qiang06Teleportation,Goebel08}. However, in the experiment
we decide to analyse the two Bell states $|\Phi^{+}\rangle$ and
$\vert\Phi^-\rangle$ to increase the efficiency - the fraction of
success - by a factor of 4. This is achieved by interfering
photons 1\&3 and photons 2\&5 on a polarizing beam splitter (PBS)
and performing a polarization analysis (PA) on the two
outputs\cite{Pan98GHZa}. With the help of a HWP, a PBS and
fibre-coupled single photon detectors, we are able to project the
input photons of the BSM onto $|\Phi^{+}\rangle$ upon the
detection of a $|+\rangle|+\rangle$ or $|-\rangle|-\rangle$
coincidence, and onto $|\Phi^{-}\rangle$ upon the detection of a
$|+\rangle|-\rangle$ or $|-\rangle|+\rangle$ coincidence (where
$|\pm\rangle=(|H\rangle \pm |V\rangle)/\sqrt{2}$). The increasing
of success efficiency allow us reducing the pump power in order to
reduce noise contributions while preserving the overall count
rate.

The projective BSMs between the data input photon 1 (2) and photon 3
(5) of the cluster state leave the remaining photons of the cluster
state 4\&6 up to a unitary transformation in the state $\vert
out\rangle_{46}$. This is the desired final state of having
performed a C-NOT operation on photons 1\&2. To demonstrate that our
teleportation-based C-NOT gate protocol works for a general unknown
polarization state of photons 1\&2, we decide to measure the truth
table of our gate. That is, we measure the output for all possible
combinations of the two-qubit input in the computational basis.
However, that is not sufficient to show the quantum characteristic
of a C-NOT gate. The remarkable feature of a C-NOT gate is its
capability of entangling two separable qubits. Thus, to fully
demonstrate the successful operation of our protocol, we furthermore
choose to perform the entangling operation:
\begin{eqnarray}
&|H\rangle_T \otimes \frac{1}{\sqrt 2}(|H\rangle_C + |V\rangle_C)
\stackrel{C-NOT}{\longrightarrow} \nonumber\\ &\frac{1}{\sqrt
2}(|H\rangle_T|H\rangle_C + |V\rangle_C|V\rangle_C) =
|\Phi^+\rangle_{TC}
\end{eqnarray}

We quantify the quality of our output state by looking at the
fidelity as defined by $F=Tr(\hat\rho\vert out\rangle\langle
out\vert)$ where $\vert out \rangle$ is the theoretically desired
final state and $\hat\rho$ is the density matrix of the experimental
output state. To analyze the operation and to experimentally measure
the fidelity of the two-qubit output, we again use PAs. Depending on
the measurement setting we use quarter wave plates (QWP) or HWP in
front of the PBS.

\begin{figure}[ptb]
\begin{center}
\includegraphics
[width=6cm]{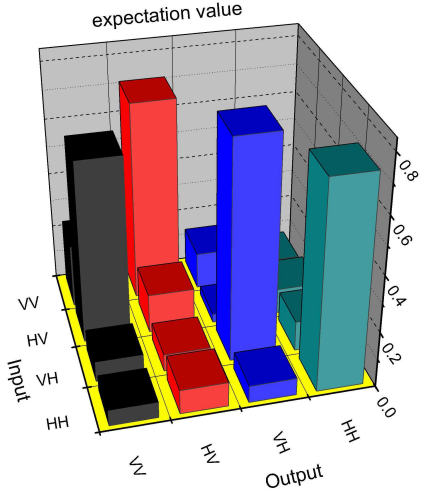}
\end{center}
\caption{Experimental results for truth table of the C-NOT gate. The
first qubit is the target and the second is the control qubit. The
average fidelity for the truth table is $0.72 \pm 0.05$.}
\label{fig3}
\end{figure}

The fidelity measurements for the truth table are straightforward.
Conditional on detecting a fourfold coincidence at the two BSMs,
we analyze the output photons 4\&6 in the computational $H/V$
basis. Depending on the type of coincidence at the BSM
($|+\rangle|+\rangle$, $|+\rangle|-\rangle$, $|-\rangle|+\rangle$,
$|-\rangle|-\rangle$), i.e. depending onto which Bell state the
photons have been projected, we analyze the output by taking into
account the corresponding unitary transformation. Since this state
analysis only involves orthogonal measurements on individual
qubits, the fidelity of the output state is directly given by the
fraction of observing the desired state. The measurement results
are shown in Fig.~\ref{fig3}. All together, 12 single-photon
detectors have been used during the whole experiment. The
experimental integration time for each possible combination of the
input photons was about 50 hours and we recorded about 120 desired
two-qubit events. The overall count rate is reduced by a factor of
1/72 due to the success probability of creating the cluster state
(1/9), the success probability of the BSMs (1/4) and due to the
loss by initializing the input state with polarizers (1/2). On the
basis of our original data, we conduct that the average fidelity
for the output states of the truth table is $0.72 \pm 0.05$.

\begin{figure}[ptb]
\begin{center}
\includegraphics
[width=5cm]{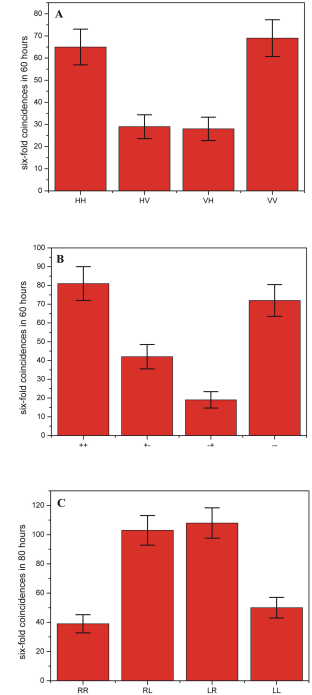}
\end{center}
\caption{Experimental results for fidelity measurement of entangled
output state. Three complementary basis are used: \textbf{(A)}
$|H\rangle/|V\rangle$ for the measurement of $\langle \hat{\sigma}_z
\hat{\sigma}_z \rangle$; \textbf{(B)} $|+\rangle/|-\rangle$ for
$\langle \hat{\sigma}_x \hat{\sigma}_x \rangle$ and \textbf{(C)}
$|L\rangle/|R\rangle = \frac{1}{\sqrt{2}} \left(|H\rangle \pm
i|V\rangle \right)$ for $\langle \hat{\sigma}_y \hat{\sigma}_y
\rangle$. The measured expectation values are: (A) $0.403 \pm 0.066$
(B) $0.462 \pm 0.057$ and (C) $-0.434 \pm 0.062$. All errors are of
statistical nature and correspond to $ \pm 1$ standard deviations.}
\label{fig4}
\end{figure}

The determination of the entangling capability is a bit more
complex. Since the output state is entangled, we are not able to
determine its fidelity by a single measurement setting. However,
with three successive local measurements on individual qubits we are
still able to accomplish our task. This can be seen by a closer look
at the fidelity under scrutiny:
\begin{eqnarray}\label{Fidelity}
F & = & Tr(\hat{\rho}|\Phi^{+}\rangle\langle\Phi^{+}|) \nonumber\\
{} & = & \frac{1}{4} Tr \left(\hat{\rho}(\hat{I} + \hat{\sigma}_{x}
\hat{\sigma}_{x} - \hat{\sigma}_{y} \hat{\sigma}_{y} +
\hat{\sigma}_{z} \hat{\sigma}_{z}) \right)
\end{eqnarray}
This implies that by measuring the expectation values
$\langle\hat{\sigma}_{x}\hat{\sigma}_{x}\rangle$,
$\langle\hat{\sigma}_{y}\hat{\sigma}_{y}\rangle$,
$\langle\hat{\sigma}_{z}\hat{\sigma}_{z}\rangle$ we can directly
obtain the fidelity of the entangled output state. The experimental
results for the correlated local measurement settings are
illustrated in Fig.~\ref{fig4}. The integration time for the first
two settings was about 60 hours and for the third setting about 80
hours. Using the above equation, we determine from our experimental
results an fidelity of $0.575 \pm 0.027$. This is well beyond the
state estimation limit of 0.40\cite{Hayashi05}. Furthermore and most
importantly, the result proofs genuine entanglement between the two
output photons, since it is above the entanglement limit of 0.50
\cite{Guehne02}.

All experimental results are calculated directly from the original
data and no noise contributions have been subtracted. The
imperfection of the fidelities is mainly due to
double-pair-emission. Furthermore, the limited interference
visibility and imperfect input states also reduce the quality of our
output states. Note that we achieve a better fidelity for the truth
table than for the entangling case. This is because for the latter
one the fidelity depends on the interference visibility at the PBS
of the BSM. All given errors are of statistical nature and
correspond to $\pm 1$ standard deviations.

Some further remarks are warranted here. With our setup we have
demonstrated in principle the feasibility of the GC scheme. Note
however, that strictly speaking we did not show complete
fault-tolerance, since in our experiment we did not encode logic
qubits onto a larger number of physical qubits. The principle of the
scheme, on the other hand, stays exactly the same and the developed
techniques of our setup can be readily extended for the case of a
larger number of encoded qubits. Along this line, the generation of
a large number of qubits, as well as an improvement of the fidelity
-- needed for realistic quantum computation -- still requires
extensive efforts in the future.

In summary, we have experimentally realized a C-NOT gate based on
quantum teleportation. With our six-photon architecture we have
experimentally demonstrated the ability to entangle two separable
qubits and have measured the truth table of the gate. This is the
first non-trivial proof-of-principle implementation of the protocol
introduced by Gottesman and Chuang. The teleportation-based scheme
offers a novel way for scalable quantum computing. Most attractively
however, this architecture allows for realizations of universal
quantum gates in a fault-tolerant manner, and in fact serves as an
important basis for measurement-based quantum computing. Thus, our
experimental demonstration represents an important step towards the
realization of resource-efficient, scalable quantum computation.

We are thankful for discussions and support on the technological
side by Tao Yang and acknowledge insightful discussions with Daniel
Gottesman. This work was supported by the Deutsche
Forschungsgemeinschaft (DFG), the Alexander von Humboldt Foundation,
the European Commission through the ERC Grant and the STREP project HIP
the National Fundamental Research Program (Grant No.2006CB921900),
the CAS, and the NNSFC. C.W. was additionally supported by the
Schlieben-Lange Program of the ESF.

\bibliography{Reference}

\section{Appendix. Teleportation-based C-NOT gate}
Here, we describe in detail the scheme of a teleportation-based
C-NOT gate. We give a step by step analyses of its implementation
with our setup, shown in Fig. 2 in the main article.

We align each $\beta$-barium borate (BBO) crystal carefully to
produce a pair of polarization entangled photons $i$ and $j$ in
the state:
\begin{equation}
|\Psi^{+}\rangle_{ij} = \frac{1}{\sqrt{2}} \left(|H\rangle_{i}
|H\rangle_{j} + |V\rangle_{i} |V\rangle_{j} \right)
\end{equation}
We use the method described in ref.~\cite{Kiesel05} to prepare the
cluster state $|\chi\rangle$. Initially, photons $3$, $4$, $5$ and
$6$ are in the state:
\begin{eqnarray}
|\Psi^{+}\rangle_{34}&\otimes&|\Psi^{+}\rangle_{56}   \nonumber
\\ =&&\frac{1}{2}
\big(|H\rangle_{3}|H\rangle_{4}|H\rangle_{5}|H\rangle_{6} +
|H\rangle_{3}|H\rangle_{4}|V\rangle_{5}|V\rangle_{6} +\nonumber \\
&&|V\rangle_{3}|V\rangle_{4}|H\rangle_{5}|H\rangle_{6} +
|V\rangle_{3}|V\rangle_{4}|V\rangle_{5}|V\rangle_{6} \big).
\end{eqnarray}
We direct photons 4 and 6 to the two input modes of a polarization
dependent beam splitter (PDBS), respectively. The transmission
$T_{H}$ $(T_{V})$ of horizontally (vertically) polarized light at
the PDBS is $1$ $(1/3)$, and we thus get
\begin{eqnarray}
& \rightarrow & \frac{1}{2}
\big(|H\rangle_{3}|H\rangle_{4'}|H\rangle_{5}|H\rangle_{6'} +
\frac{1}{\sqrt{3}}|H\rangle_{3}|H\rangle_{4'}|V\rangle_{5}|V\rangle_{6'} \nonumber \\
& & +\frac{1}{\sqrt{3}}
|V\rangle_{3}|V\rangle_{4'}|H\rangle_{5}|H\rangle_{6'}\nonumber \\
&
&-\frac{1}{3}|V\rangle_{3}|V\rangle_{4'}|V\rangle_{5}|V\rangle_{6'}\big).
\end{eqnarray}
Here we have neglected terms with more than one photon in a single
output mode of the PDBS, since in the experiment we post select
only terms that lead to a six-fold coincidence.

In order to symmetrize the state we place a PDBS' $\left(T_{H} =
1/3, T_{V} = 1\right)$ in each output mode of the PDBS and receive
\begin{eqnarray}
& \rightarrow & \frac{1}{6}
\big(|H\rangle_{3}|H\rangle_{4''}|H\rangle_{5}|H\rangle_{6''} +
|H\rangle_{3}|H\rangle_{4''}|V\rangle_{5}|V\rangle_{6''} \nonumber \\
& & + |V\rangle_{3}|V\rangle_{4''}|H\rangle_{5}|H\rangle_{6''} -
|V\rangle_{3}|V\rangle_{4''}|V\rangle_{5}|V\rangle_{6''}\big).
\end{eqnarray}
This is already the desired four-qubit cluster state up to local
unitary operations. To bring it to the desired form, we place
half-wave plates (HWPs) -- with an angle of $22.5^{\circ}$ between
the fast and the horizontal axis -- into arms 3 and 4. This yields
\begin{eqnarray}
\rightarrow  &&\left(|H\rangle_{3}|H\rangle_{4''} +
|V\rangle_{3}|V\rangle_{4''} \right) |H\rangle_{5}|H\rangle_{6''} \nonumber \\
&& + \left(|H\rangle_{3}|V\rangle_{4''} +
|V\rangle_{3}|H\rangle_{4''} \right) |V\rangle_{5}|V\rangle_{6''}
 \nonumber \\ &=& |\chi\rangle_{34''56''},
\end{eqnarray}
where we have neglected the overall pre-factor 1/6 and we arrive
at the desired ancillary four-photon cluster state $|\chi\rangle$
described in ref.~\cite{Gottesman99}.

Photons 1 and 2 constitute the input to our C-NOT gate. We assume
that they are in a most general input state
$|\Psi_{in}\rangle_{12}$, where:
\begin{eqnarray}
|\Psi_{in}\rangle_{ij} &=& \alpha |H\rangle_{i} |H\rangle_{j} +
\beta |H\rangle_{i} |V\rangle_{j}   \nonumber\\
&&+\gamma |V\rangle_{i} |H\rangle_{j} + \delta |V\rangle_{i}
|V\rangle_{j}
\end{eqnarray}
The pre-factors $\alpha$, $\beta$, $\gamma$ and $\delta$ are four
arbitrary complex numbers satisfying $|\alpha|^{2} + |\beta|^{2} +
|\gamma|^{2} + |\delta|^{2} = 1$. Before we proceed, let us define
the desired output state after a C-NOT operation:
\begin{eqnarray}
|\Psi_{out}\rangle_{ij}  & = &  U^{C-NOT}|\Psi_{in}\rangle_{ij} \nonumber\\
& = & \alpha |H\rangle_{i} |H\rangle_{j} + \beta |V\rangle_{i}
|V\rangle_{j} + \nonumber\\
&&\gamma |V\rangle_{i} |H\rangle_{j} + \delta |H\rangle_{i}
|V\rangle_{j}
\end{eqnarray}
The target qubit $i$ is flipped on the condition that the control
qubit $j$ is in the state $|V\rangle$.

We can now express the combined state of all six photons in terms
of Bell states for photons 1\&3 and 2\&5 and in terms of the
desired output state $|\Psi_{out}\rangle_{46}$ for photons 4\&6
with corresponding Pauli operations:
\begin{widetext}
\begin{align}
\begin{array}{rrcr}
|\Psi_{in}\rangle_{12}\otimes|\chi\rangle_{3456}  = &&&\\
|\Phi^{+}\rangle_{13}|\Phi^{+}\rangle_{25} &
|\Psi_{out}\rangle_{46} & +
|\Phi^{+}\rangle_{13}|\Phi^{-}\rangle_{25} & \hat{\sigma}_{z}^{6}
|\Psi_{out}\rangle_{46} \\
+ |\Phi^{+}\rangle_{13}|\Psi^{+}\rangle_{25} &
\hat{\sigma}_{x}^{4}\hat{\sigma}_{x}^{6}|\Psi_{out}\rangle_{46} &
+ |\Phi^{+}\rangle_{13}|\Psi^{-}\rangle_{25} &
\hat{\sigma}_{x}^{4}\hat{\sigma}_{x}^{6}\hat{\sigma}_{z}^{6}
|\Psi_{out}\rangle_{46} \\
+ |\Phi^{-}\rangle_{13}|\Phi^{+}\rangle_{25} &
\hat{\sigma}_{z}^{4}\hat{\sigma}_{z}^{6}|\Psi_{out}\rangle_{46} &
+ |\Phi^{-}\rangle_{13}|\Phi^{-}\rangle_{25} &
\hat{\sigma}_{z}^{4}
|\Psi_{out}\rangle_{46} \\
+ |\Phi^{-}\rangle_{13}|\Psi^{+}\rangle_{25} &
\hat{\sigma}_{x}^{4}\hat{\sigma}_{z}^{4}\hat{\sigma}_{x}^{6}\hat{\sigma}_{z}^{6}
|\Psi_{out}\rangle_{46} & +
|\Phi^{-}\rangle_{13}|\Psi^{-}\rangle_{25} &
\hat{\sigma}_{x}^{4}\hat{\sigma}_{z}^{4}\hat{\sigma}_{x}^{6}
|\Psi_{out}\rangle_{46}  \\
+ |\Psi^{+}\rangle_{13}|\Phi^{+}\rangle_{25} &
\hat{\sigma}_{x}^{4}|\Psi_{out}\rangle_{46} & +
|\Psi^{+}\rangle_{13}|\Phi^{-}\rangle_{25} &
\hat{\sigma}_{x}^{4}\hat{\sigma}_{z}^{6}
|\Psi_{out}\rangle_{46} \\
+ |\Psi^{+}\rangle_{13}|\Psi^{+}\rangle_{25} &
\hat{\sigma}_{x}^{6}|\Psi_{out}\rangle_{46} & +
|\Psi^{+}\rangle_{13}|\Psi^{-}\rangle_{25} &
\hat{\sigma}_{x}^{6}\hat{\sigma}_{z}^{6}
|\Psi_{out}\rangle_{46}  \\
+ |\Psi^{-}\rangle_{13}|\Phi^{+}\rangle_{25} &
\hat{\sigma}_{x}^{4}\hat{\sigma}_{z}^{4}\hat{\sigma}_{z}^{6}|\Psi_{out}\rangle_{46}
& + |\Psi^{-}\rangle_{13}|\Phi^{-}\rangle_{25} &
\hat{\sigma}_{x}^{4}\hat{\sigma}_{z}^{4}
|\Psi_{out}\rangle_{46} \\
+ |\Psi^{-}\rangle_{13}|\Psi^{+}\rangle_{25} &
\hat{\sigma}_{z}^{4}\hat{\sigma}_{x}^{6}\hat{\sigma}_{z}^{6}|\Psi_{out}\rangle_{46}
& + |\Psi^{-}\rangle_{13}|\Psi^{-}\rangle_{25} &
\hat{\sigma}_{z}^{4}\hat{\sigma}_{x}^{6}
|\Psi_{out}\rangle_{46}  \\
\end{array}
\end{align}
\end{widetext}
With the help of polarizing beam splitters, in our experiment we
are able to identify the Bell states $|\Phi^{\pm}\rangle_{13}$ and
$|\Phi^{\pm}\rangle_{25}$, i.e. we project the combined state of
photons $1$, $2$, $3$ and $5$ onto one of the four possibilities
$|\Phi^{\pm}\rangle_{13}|\Phi^{\pm}\rangle_{25}$. We thus have to
consider four different results of the BSMs: \vskip 1.0cm
\begin{center}
\begin{tabular}{|c|r|}
\hline
Result of BSMs    &   Output state  \\
    \hline
$|\Phi^{+}\rangle_{13}|\Phi^{+}\rangle_{25}$ & $|\Psi_{out}\rangle_{46}$\\
$|\Phi^{+}\rangle_{13}|\Phi^{-}\rangle_{25}$ & $\hat{\sigma}^{6}_{z}|\Psi_{out}\rangle_{46}$\\
$|\Phi^{-}\rangle_{13}|\Phi^{+}\rangle_{25}$ & $\hat{\sigma}^{4}_{z}\hat{\sigma}^{6}_{z}|\Psi_{out}\rangle_{46}$\\
$|\Phi^{-}\rangle_{13}|\Phi^{-}\rangle_{25}$ & $\hat{\sigma}^{4}_{z}|\Psi_{out}\rangle_{46}$\\
    \hline
\end{tabular}
\end{center}
\vskip 1.0cm To receive the desired final state of photons $4$ and
$6$, we have to apply corresponding Pauli operations, depending on
the outcome of the BSMs.

\end{document}